\documentclass[12pt,titlepage]{article}
\usepackage{apacite}
\usepackage{lscape}
\usepackage{graphicx}
\usepackage{lmodern}
\usepackage{booktabs, tabularx,amsmath}
 
\usepackage{caption}

\usepackage{fancyhdr}   
\fancypagestyle{plain}{\fanyhf{}}      
\advance \topmargin by -0.7 cm
\advance \oddsidemargin by -1.6 cm
\renewcommand{\baselinestretch}{1.5}
\begin{document}
\begin{flushleft}
\pagestyle{myheadings}
\setlength{\parindent}{1.5 cm}
\setcounter{page}{1}
\leftline{Running head: Order Constraints}
\vfil
\centerline{Parametric Order Constraints in Multinomial Processing Tree Models:} \par
\centerline{An Extension of Knapp and Batchelder (2004)}
\bigskip
\centerline{Karl Christoph Klauer, Henrik Singmann, and David Kellen}
\centerline{Albert-Ludwigs-Universit\"at Freiburg}

\vfil

\centerline{Author Note}

Karl Christoph Klauer, Institut f\"ur Psychologie;
Henrik Singmann, Institut f\"ur Psychologie;
David Kellen, Institut f\"ur Psychologie.

The first two authors contributed to equal extents to the manuscript.

Correspondence concerning this article should be addressed to K.
C. Klauer at the Institut f\"ur Psychologie, Albert-Ludwigs-Universit\"at
Freiburg, \hbox{D-79085 Freiburg}, Germany. Electronic
mail may be sent to
\hbox{christoph.klauer@psychologie.uni-freiburg.de}.
\par\noindent

\vfil
\noindent
Address information for Christoph Klauer (corresponding author):
\begin{itemize}
\item E-mail: \hbox{christoph.klauer@psychologie.uni-freiburg.de}
\item Phone: +49 761 2032469
\item Postal address:
\begin{itemize}
	\item Institut f\"ur Psychologie
	\item Albert-Ludwigs-Universit\"at Freiburg
	\item D-79085 Freiburg
	\item Germany
\end{itemize}
\end{itemize}

\newpage

\centerline{\bf Abstract}

\noindent
Multinomial processing tree (MPT) models are tools for disentangling the contributions 
of latent cognitive processes in
a given experimental paradigm. The present note analyzes MPT models subject to order constraints
on subsets of its parameters. The constraints that we consider
frequently arise  in cases where the response categories are ordered in some sense such as in 
confidence-rating data, Likert scale data, where
graded guessing tendencies or response biases 
are created via base-rate or payoff manipulations, in the analysis of contingency tables
with order constraints, and in many other cases. We show how to
construct an MPT model without order constraints that is statistically equivalent to
the MPT model with order constraints. This new closure result extends the mathematical analysis of the
MPT class, and it offers an approach to order-restricted inference 
that extends the approaches discussed by Knapp and Batchelder (2004). The usefulness of
the method is illustrated by means of an analysis of an order-constrained version of the 
two-high-threshold model for confidence ratings.

\vskip 2 true cm
\noindent
KEYWORDS: Multinomial processing tree models, mathematical models, categorical data, multinomial distribution

\newpage

Multinomial processing tree (MPT) models are used to measure cognitive processes in many    
areas of psychology (for reviews, see Batchelder \& Riefer, 1999; 
Erdfelder et al., 2009). They are models for categorical data. MPT models are typically tailored to
a given experimental paradigm and specify how the most important
processes assumed to be involved in data generation in the paradigm interact
to produce observable responses.

As an example consider the two-high-threshold model (2HTM; Snodgrass \& Corwin, 1988).
 The 2HTM  is tailored to memory experiments in which old/new judgments
are requested for previously studied items intermixed with new items. In many such experiments,
participants are also asked to rate their confidence in each
``old'' or ``new'' judgment. Figure 1 shows a version
of the model for a confidence rating scale with three points, labeled
``high'', ``medium'', and ``low'' (Br\"oder, Kellen, Sch\"utz, \& Rohrmeier, 2013; Klauer \& Kellen, 2011). 
Responses are mediated via three
latent states, labeled ``detect old'', ``detect new'', and ``no detection''.

Parameters $D_o$ and $D_n$  define a stimulus-state mapping.
$D_o$ is the probability of entering the ``detect old'' state
for an old item; $D_n$ of entering the ``detect new'' state
for a new item; the ``no detection'' state is entered with probability $1-D_o$ and
$1-D_n$ for old and new items, respectively. The remaining parameters
 define state-response mappings. Given one of the two
``detect'' states, the old/new judgment is invariably correct as regards the old versus new status of the
test item, and the
parameters $s_l$, $s_m$, and $s_h$ ($s_l+s_m+s_h=1$) quantify 
the probabilities of selecting, in order, the low, medium, and high confidence level in the old/new response.$^1$
In the absence of detection, there is a guessing bias captured by probability parameter $g$,
quantifying the probability of guessing ``old'' rather than ``new''.
 Given that ``old'' is guessed, parameters $o_l,$ $o_m$, and $o_h$ 
with $o_l+o_m+o_h=1$ parameterize the probabilities for the three confidence levels;
given that ``new'' is guessed, $n_l$, $n_m$, and $n_h$ parameterize these probabilities. 

As can be seen, MPT models assume that 
observed category counts arise from processing
branches consisting of separate conditional links or stages. Each branch probability
is the product of its conditional link probabilities, and more than one branch can
terminate in the same observed category (Hu \& Batchelder, 1994).

In most cases, the models are eventually 
represented as so-called binary MPT models (Purdy \& Batchelder, 2009), because many software tools for analyzing MPT models
require binary MPT models as input.
In a binary MPT model, exactly two links go out from each non-terminal
node. The two links are labeled by two parameters that sum to one. 
One of these is redundant and is replaced by one minus
the other parameter so that the remaining model parameters are functionally independent, each such
parameter ranging from 0 to 1. It is straightforward to transform a non-binary MPT model into
 a statistically equivalent binary MPT model (Hu \& Batchelder, 1994). Two models are
statistically equivalent if they can predict the same sets of response probabilities.

In applications, it is not uncommon that order constraints are predicted to hold for subsets of
the functionally independent parameters of binary MPT models (Baldi \& Batchelder, 2003; Knapp \& Batchelder, 2004), 
and Knapp and Batchelder have shown that the model class is closed under
one or more non-overlapping linear orders of parametric constraints. That is,
a new non-constrained binary MPT model can be constructed using a different set of 
functionally independent parameters that is statistically equivalent to
the original model with the order constraints. 

Here, we consider a different set of order constraints that regularly
arise in applications and that are not covered by Knapp and Batchelder (2004).
The order constraints frequently arise where response categories are ordered in some sense such as for confidence ratings 
or Likert scales. They also arise where participants discriminate
between two or more categories of items and the probabilities of guessing the
categories in ``no detection'' states can be assumed to be ordered; for example, because base rates or payoffs systematically 
differ between the categories. Our results also apply to the important case of order constraints on the
probabilities of a multinomial or product-multinomial distribution that is frequently encountered within
and outside psychology (e.g., Agresti \& Coull, 2002).
We show that an MPT model with the order constraints can 
be represented in the form of a statistically equivalent non-constrained MPT model. This new closure
property contributes to the structural analysis of the MPT class and is
immediately useful for analyzing cases in which the order constraints are
to be imposed upon the parameters as exemplified below.

Considering, for example, the 2HTM  for confidence ratings, 
a psychologically plausible constraint on the parameters
for the confidence levels in the ``no detection'' state is that the preference
for a given confidence level should decline from
lowest to highest confidence levels, reflecting the respondent's uncertainty in the
absence of detection: $o_l \ge o_m \ge o_h$ and
$n_l \ge  n_m \ge  n_h$. Conversely, in ``detect'' states, 
 the preference for a given confidence level should increase from lowest to highest confidence, at least for scales with only
a few confidence levels: $s_l \le  s_m \le  s_h$.

Imposing such constraints sharpens the distinctions between ``no detection'' and ``detect'' states
by highlighting plausible qualitative differences between them. 
When satisfied by the underlying probability distribution, 
the constraints  contribute to making the estimation of the parameters $D_n$ and $D_o$ of the
stimulus-state mapping more precise, focused,  and robust, and they considerably increase the model's 
parsimony as elaborated on below.

These order constraints are imposed on functionally dependent parameters (e.g., $s_l$, $s_m$, and $s_h$ have to
sum to 1 and are therefore not independent). Hence, they are not covered by Knapp and Batchelder's (2004) approach
to order constraints for independent parameters. Nevertheless, it 
is possible to express them in the language of MPT models. 

The next section  describes how to transform MPT models with order constraints of this kind
into equivalent non-constrained MPT models.$^2$
Finally, we illustrate the new method by comparing versions of the 2HTM  with and without
order constraints in terms of model complexity and in terms of their description of 
a dataset by Koen and Yonelinas (2010). The general discussion expands on the
advantages of the new method for estimation and inference with order-constrained models.

\centerline{\bf Order Constraints on Multinomial Probabilities}

We consider a basic subtree with
a root, no other non-terminal node, and two or more links going out from its root as shown on the left side
of Figure 2. This subtree might occur at one or more places in the
tree representation of the complete model. With regard to the overall MPT model, its terminal nodes $A_1, \ldots, A_k$
represent either other subtrees or
observable categories. 
For example, the subtree with three links labeled by
parameters $s_l$, $s_m$, and $s_h$ in the above 2HTM  occurs at
two places in the processing tree representation (see Figure 1).

Our basic result describes how to represent a linear order $\eta_1\ge \eta_2 \ge \dots \ge \eta_k$
on the parameters of the subtree by means of a statistically equivalent MPT model without order constraints on the parameters. 
As shown in Figure 2 the solution is to replace each occurrence of the subtree 
in question by  the subtree on the right side in Figure 2. Replacement means
that the tree on the right side replaces the subtree on the left
side wherever it occurs in the processing-tree representation. 
Furthermore, whatever is appended at the terminal node $A_j$
of an occurrence of the original subtree is appended in the replacing subtree wherever
a terminal node labeled $A_j$ occurs.
As for the reparameterizations in Knapp and Batchelder (2004) this regularly
implies an increase in the size of the processing tree. 

We prove the following two theorems:

{\bf Theorem 1:} For the tree shown on the right side of Figure 2 and any set of non-negative
parameters $\lambda_i$, $i=1,\ldots, k$, with $\sum_{i=1}^k\lambda_i=1$,
the probabilities of outcomes $A_i$ are ordered as $P(A_1)\ge P(A_2) \ge \ldots \ge
P(A_k)$.

Theorem 1 states that the tree indeed imposes an order constraint
on its $k$ outcome probabilities. Theorem 2 complements this by showing that any ordered
set of  probabilities can be represented in this form:

{\bf Theorem 2:} For any set of ordered probabilities $\eta_i$ with $\eta_1\ge
\eta_2 \ge \ldots \ge \eta_k$, $\sum_i \eta_i =1$, there exist non-negative
values $\lambda_i$, $i=1,\ldots,k$ with $\sum_i \lambda_i =1$ such that
the outcome probabilities of the tree on the right side of Figure 2 are given
by $P(A_i)=\eta_i$, $i=1,\ldots,k$.

{\bf Proof of Theorem 1.}
Note that the outcome probabilities of the tree on the right side of Figure 2 are mixtures with mixture coefficients $\lambda_j$. 
The first mixture component is given by the probability distribution with $P(A_1)=1$ and $P(A_j)=0$, $j>1$;
the second by $P(A_1)=P(A_2)=1/2$ and $P(A_j)=0$, $j>2$; the last by $P(A_j)=1/k$ for all $j$.
Each mixture component satisfies the inequalities, $P(A_j)\ge P(A_{j+1})$, $j=1,\ldots,k-1$.
This completes the proof.

{\bf Proof of Theorem 2.}
Define mixture coefficients $\lambda_j$ as follows:
\begin{equation}
 \lambda_k=k \eta_k \mbox{ and } \lambda_j= j(\eta_{j} - \eta_{j+1}), j=1,\ldots,k-1.  
\end{equation}
By the premises of Theorem 2, it is immediate that $\lambda_j\ge0$.
It is furthermore
easy to see via simple manipulations that $\sum_{j=1}^k \lambda_j =\sum_{j=1}^k\eta_j=1$ and that
$\eta_j= \sum_{i=j}^k \frac{\lambda_i}{i}$.  Hence, $P(A_j)=\sum_{i=j}^k \frac{\lambda_i}{i}=\eta_j$, $j=k,k-1,\ldots,1$.
This completes the proof.

\leftline{\bf Extensions}

In this section, we consider linear-order constraints on only a subset of the $\eta_i$,
non-overlapping linear-order constraints on the $\eta_i$,  more general partial orders,
and all such orders for $k=3$ and $k=4$.

{\bf Linear orders on subsets and multiple non-overlapping linear orders.}
Theorems 1 and 2 cover the case of a total order on all $k$ parameters $\eta_i$. It is straightforward to
extend the results to the case of a  partial linear order on only a subset of the $\eta_i$. For example,
assume $k=6$ and that the constraint $\eta_1 \ge \eta_2 \ge \eta_3$ is to be imposed. It is easy to see
that this translates into the constraint $\theta_1\ge \theta_2\ge \theta_3$ in the reparameterization depicted
in Figure 3, where $\theta_1+\theta_2+\theta_3=1$ so that Theorems 1 and 2 apply.

The reparameterization in Figure 3 also shows how two or more non-overlapping linear-order constraints can be imposed
upon the $\eta_i$ with $\sum_i \eta_i =1$. For example, if in addition $\eta_4 \ge \eta_5 \ge \eta_6$ is to be 
imposed, this could be implemented by imposing the constraint $\theta_4 \ge \theta_5 \ge \theta_6$, where
$\theta_4+\theta_5+\theta_6=1$, in terms of the parameters $\theta_i$ in Figure 3.

{\bf General partial orders.} More general partial orders can often be treated by the following idea. The probability 
distributions ${\boldsymbol \eta} = (\eta_1,\ldots,\eta_k)$ that satisfy a set of 
linear
inequality constraints form a convex polytope. Specifically, they can exhaustively 
be represented as  mixtures
of certain fixed probability distributions  
${\boldsymbol \eta}_1$, ${\boldsymbol \eta}_2$, \ldots, ${\boldsymbol \eta}_l$ that we refer to as vertices.
For small problems, the vertices can be found graphically; in complex cases, linear programming
algorithms can be used.

For example, for the linear order with $\eta_1\ge \eta_2 \ge \ldots \ge \eta_k$, the vertices
are given by the above-mentioned $k$ mixture components, ${\boldsymbol \eta}_1=(1,0,\ldots,0)$, 
${\boldsymbol \eta}_2 = (1/2,1/2,0,\ldots,0)$, $\ldots$, ${\boldsymbol \eta}_k=(1/k,\ldots,1/k)$,
and the model parameters $\lambda_j$ of the non-constrained MPT model expressing the order constraint 
are simply the mixture weights. The subtree following  $\lambda_j$ codes the probability
distribution specified in vertex ${\boldsymbol \eta}_j$

{\bf Partial orders with {\boldmath $k=3$.}} This immediately solves the case of orderings among $k$ parameters that can be represented by
$k$ vertices. Consider, for example, $k=3$, and the ordering $\eta_1\le \eta_2$ and $\eta_1\le \eta_3$.
This defines a convex polytope with the three vertices ${\boldsymbol \eta}_1=(0,0,1)$, 
${\boldsymbol \eta}_2 = (0,1,0)$, and ${\boldsymbol \eta}_3 = (1/3,1/3,1/3)$. 
Hence, a non-constrained MPT model representing the order restrictions has three mixture coefficients $\lambda_i$, $i=1,\ldots,3$,
as parameters. Each $\lambda_i$ is linked to a subtree coding the respective probability distributions ${\boldsymbol \eta}_i$.

In other cases, more than $k$ vertices are required. For example, the ordering $\eta_1\ge\eta_2$,
$\eta_1\ge\eta_3$ defines a polytope with four vertices, ${\boldsymbol \eta}_1=(1,0,0)$, 
${\boldsymbol \eta}_2 = (1/2,1/2,0)$, ${\boldsymbol \eta}_3 = (1/2,0,1/2)$, and ${\boldsymbol \eta}_4 = (1/3,1/3,1/3)$.
Using four mixture coefficients $\lambda_i$ to represent the order-constrained MPT model
as a non-constrained MPT model is possible, but results in an overparameterized model. This means
that certain analyses and inferences available for MPT models without overparameterization cannot be done. 
For example, although it is possible to  determine the model's maximum
log-likelihood and $G^2$ goodness-of-fit statistic and to bootstrap its distribution (e.g., Singmann \& Kellen, 2013), 
it will not be possible to determine unique parameter estimates. 

For the present example, it can however be shown that representing the four mixture coefficients $\lambda_i$
by two independent parameters $\theta_1$ and $\theta_2$ with $0\le \theta_i\le 1$
such that $\lambda_1=(1-\theta_1)(1-\theta_2)$, $\lambda_2=\theta_1(1-\theta_2)$, $\lambda_3=(1-\theta_1)\theta_2$,
and $\lambda_4=\theta_1\theta_2$ is sufficient to span the entire polytope defined by the above vertices (see appendix). This immediately
leads to a non-redundant parameterization using two independent parameters $\theta_1$ and
$\theta_2$. 

But an analogous construction does not in general guarantee this in other cases. Nevertheless, for most purposes
it is sufficient that a non-redundant parameterization is found that covers the point of maximum likelihood
 (as determined, for example, via estimating the overparameterized model in a first step)
and its local environment. This is usually not difficult to achieve departing from the vertex representation.

{\bf Partial orders with {\boldmath $k=4$.}} Tables 1 and 2 show the vertex representations of the possible partial orders involving four outcomes A, B, C, and D
as shown in Figure 4. The online supplemental material sketches a heuristic for determining non-redundant
parameterizations of the mixture coefficients in these and more complex cases.

\leftline{\bf Application: Order Constraints on the 2HTM}

As already noted, the 2HTM distinguishes ``detect'' states and a ``no detection'' state. These are
latent states, and a state-response mapping is required to link the states to the observable
responses. Klauer and Kellen (2011) and Br\"oder et al. (2013) proposed relatively unrestricted
state-response mapping in which it is possible, for example, that extreme confidence ratings
would be preferred in a ``no detection'' state and that low confidence ratings
would be preferred in a ``detect'' state. This has prompted criticisms to the effect that the models deal
with rating scales in an arbitrary and post-hoc manner (e.g., Dube, Rotello, \& Pazzaglia, 2013, 
Pazziaglia, Dube, \& Rotello, 2013; see also Batchelder \& Alexander, 2013). According to these critics,
the model is overly complex (Dube, Rotello, \& Heit, 2011).

Using the present results, it is possible to define an order-constrained 2HTM for confidence
ratings, 2HTM$_r$, that maps the ``no detection'' state so that cautious low confidence ratings receive
most probability mass and more extreme confidence ratings successively less mass. Conversely, for detect
states, the preference for confidence levels increases from low confidence to high confidence levels
in 2HTM$_r$.
Specifically,
2HTM$_r$ imposes the constraints that $s_h \ge s_m \ge s_l$  for the mapping from ``detect'' states to responses,
and that $o_l \ge o_m \ge o_h$
and $n_l\ge n_m\ge n_o$ for  the mapping of the ``no detection'' state to responses (see Figure 1 for the 2HTM).
These constraints remove most of the
less plausible state-response mappings that are admissible under the original 2HTM for
confidence ratings. At the
same time, they strongly curtail the mathematical flexibility of the model.

Because these constraints can now be implemented within the MPT framework, we can use
standard MPT software to estimate, test, and analyze the model. For example, 
we used the R package MPTinR (Singmann \& Kellen,
2013) to quantify model flexibility of the 2HTM and the 2HTM$_r$
based on the minimum-description-length principle (Gr\"unwald, 2007) that takes
flexibility due to a model's functional form into account, including constraints on flexibility 
due to inequality constraints.
In the present context, one representation of the minimum-description-length principle is given by
the Fisher information approximation (FIA) index (but see Heck, Moshagen, \& Erdfelder, 2014).
FIA is a  model selection
index and  describes model flexibility
by a  penalty that is added to a likelihood measure of the model's (mis)fit.
The model with the smallest index value is preferred. Wu, Myung, and Batchelder (2010a, 2010b)
developed methods to compute FIA for binary MPT models
that is implemented in MPTinR.

The penalty term for model flexibility in FIA comprises two additive terms. One term depends upon the
number of parameters and the sample size similar to the penalty term in BIC. A second term quantifies 
model flexibility due to functional form, and 2HTM
and 2HTM$_r$ differ in the size of this penalty term. Using the above reparameterization
and MPTinR, we find the penalties due to flexibility to be -0.01 and -5.22 for 2HTM and 2HTM$_r$, respectively.
Because FIA and the normalized maximum-likelihood index operate on a log-likelihood scale,
this means that the loss in  
goodness of fit in terms of $G^2$ (Batchelder \& Riefer, 1999) must be larger than
10.42 $=2[-0.01-(-5.22)]$ before the parsimonious 2HTM$_r$ should be abandoned in favor of the 
non-constrained 2HTM. This demonstrates that the constraints on flexibility
imposed by the order constraints are substantial.

For example, reanalyzing data from Koen and Yonelinas (2010, pure condition), the 2HTM without order constraints
achieves a $G^2$ statistic of 3.35 and a FIA index of $8.28$ (up to an additive constant).
The 2HTM with order constraints achieves a $G^2$ statistic of 9.95 and a FIA index of $6.37$.
Thus, the loss in goodness of fit is outweighed by the parsimony of the constrained model, and
the 2HTM$_r$ should be preferred.

\centerline{\bf Discussion}

In this note, we extended Knapp and Batchelder's (2004) approach to order constraints in MPT models 
to the case of order restrictions on non-independent parameters constrained to sum to one. These
restrictions frequently arise in cases where response outcomes themselves are ordered in some sense such as in 
confidence-rating data, Likert scale data (B\"ockenholt, 2012; Klauer \& Kellen, 2011) or where
graded guessing tendencies or response biases are created via base-rate or payoff manipulations.
Alternatively, the restrictions can directly arise from theoretical predictions (e.g., 
Ragni, Singmann, \& Steinlein, 2014).

The results are useful because they make available the growing toolbox for
statistical analyses of MPT model for the analysis of order-constrained MPT models. 
As already exemplified, the toolbox comprises
algorithms and software for the computation of FIA (Moshagen, 2010; Singmann \& Kellen, 2013; Wu et al., 2010a, 2010b),
but also a number of software tools to estimate and fit the models (see Klauer, Stahl, \& Voss, 2012
for a review), algorithms for Bayesian hierarchical model extensions that capitalize on the
model structure (Klauer, 2010; Matzke, Dolan, Batchelder, \& Wagenmakers, in press; Smith \& Batchelder, 2010), algorithms and software for hierarchical
latent-class extensions of MPT models in a classical inferential framework (Klauer, 2006, Stahl \& Klauer, 2007), 
and algorithms for computing Bayes factors between competing MPT models (Vandekerckhove, Matzke, \& Wagenmakers, in press).

The present results also apply to the important case of order constraints on the probabilities of an observable multinomial
or product-multinomial distribution, a case with many occurrences within and outside psychology --- for example, in the analysis
of contingency tables with order constraints (Agresti \& Coull, 2002). The case is treated by Robertson, Wright, and Dykstra (1988, chap. 5) 
via their elegant isotonic regression method. Note, however, that expressing order structures on 
observable category probabilities in the MPT framework has the advantage 
of making available the above-mentioned tools for estimation and inference in classical and Bayesian frameworks.
As one example, using MPTinR (Singmann \& Kellen, 2013), the distribution of goodness-of-fit statistic $G^2$ can be assessed
via bootstrap methods under the null hypothesis that the constraints are truly in force,
side-stepping the numerically difficult task to evaluate its asymptotic so-called $\bar\chi^2$ distribution.

On the theoretical side, these results contribute to the study of the class of MPT models. They
show that the class is closed under this further set of order constraints. This flexibility
is surprising given that most other classes of statistical models that we are aware of would
not be invariant under transformations as in Equation 1, 
nor capable of expressing distributions with and without
the present kind of order constraints within the same class of models. The ability 
of the model class to encompass the present and other kinds of meaningful order constraints adds to its usefulness.

\newpage

\centerline{\bf Appendix}

We show that each probability distribution on three outcomes
with $\eta_1\ge \eta_2$ and $\eta_1\ge \eta_3$ and $\sum_{i=1}^3\eta_i =1$
can be represented as a mixture with independent parameters $\theta_1$, $\theta_2$,
$0\le \theta_i \le 1$ as follows:

$$
\left(
\begin{array}{c}
\eta_1 \\
\eta_2 \\
\eta_3 \\
\end{array} \right)
=(1-\theta_1)(1-\theta_2)
\left(
\begin{array}{c}
1 \\
0 \\
0 \\
\end{array} \right)
+\theta_1(1-\theta_2)
\left(
\begin{array}{c}
1/2 \\
1/2 \\
0 \\
\end{array} \right)
+ (1-\theta_1)\theta_2
\left(
\begin{array}{c}
1/2 \\
0 \\
1/2 \\
\end{array} \right)
+\theta_1\theta_2
\left(
\begin{array}{c}
1/3 \\
1/3 \\
1/3 \\
\end{array} \right).
$$
The proof proceeds by showing that this equation can be solved for $\theta_1$ and $\theta_2$ given $\eta_1$, $\eta_2$, and $\eta_3$.
Note that $\eta_1-\eta_2-\eta_3 = (1-\theta_1)(1-\theta_2)-\frac{1}{3}\theta_1\theta_2$.
Setting $\Delta=\eta_3-\eta_2$, it follows that $\Delta = \frac{1}{2}(\theta_2-\theta_1)$, hence $\theta_2 = 2\Delta + \theta_1$.
Substituting this in the previous equation yields: $2\eta_1-1=\eta_1-\eta_2-\eta_3=\frac{1}{3}(3-6\Delta-2\theta_1(3-2\Delta)+2\theta_1^2)$.
Some manipulations show that this is equivalent to $[\theta_1- \frac{1}{2}(3-2\Delta)]^2=[\frac{1}{2}(3-2\Delta)]^2+3(\eta_1+\Delta-1)$.

For this quadratic equation to be solvable in $\theta_1$, it needs to be shown that $[\frac{1}{2}(3-2\Delta)]^2+3(\eta_1+\Delta-1)\ge 0$.
This is equivalent to $[\frac{1}{2} + (1-\Delta)]^2\ge 3(1-\eta_1-\Delta)$. Because $\eta_1+\eta_2+\eta_3=1$, this is equivalent to
$[\frac{1}{2} + \eta_1+2\eta_2]^2\ge 6\eta_2$. Because $\eta_1\ge\eta_2$, this is true if $[\frac{1}{2} + 3\eta_2]^2\ge 6\eta_2$.
This is equivalent to $(3\eta_2-\frac{1}{2})^2\ge 0$.

Hence, one solution of the above equation is $\theta_1 = \frac{1}{2}(3-2\Delta)-\sqrt{[\frac{1}{2}(3-2\Delta)]^2-6\eta_2}$. We will
show that $0\le \theta_1\le 1$. $\theta_1\ge 0$ is to see noting that $\frac{1}{2}(3-2\Delta)=\frac{1}{2}+\eta_1+2\eta_2\ge 0$.
Furthermore, because of this, $\theta_1\le 1$ is equivalent to $\frac{1}{2}+\eta_1+2\eta_2-\sqrt{(\frac{1}{2}+\eta_1+2\eta_2)^2-6\eta_2} \le 1$ or to
$\eta_1+2\eta_2-\frac{1}{2} \le \sqrt{(\frac{1}{2}+\eta_1+2\eta_2)^2-6\eta_2}$. 
This is trivially true if the term to the left,  $\eta_1+2\eta_2-\frac{1}{2}$, is smaller than zero. If it is non-negative, on the other hand,
this is equivalent to $(\eta_1+2\eta_2-\frac{1}{2})^2\le (\frac{1}{2}+\eta_1+2\eta_2)^2-6\eta_2$, which is equivalent to $\eta_1\ge\eta_2$ as simple
manipulations show. Because the equations are symmetrical in $\theta_1$ and $\theta_2$, interchanging the roles of $\theta_1$ and $\theta_2$
and those of $\eta_2$ and $\eta_3$ shows that $\theta_2=\theta_1 + 2\Delta$ also ranges between 0 and 1. This completes the proof.

\newpage

\centerline{\bf Footnotes}

\smallskip\noindent
$^1${There are well-documented  
response-style
effects such as preferring moderate over extreme responses or vice versa
as moderated by contextual and personality factors (B\"ockenholt, 2012). In the light of these effects,
it is reasonable to assume that ``detect'' states are not necessarily always
mapped on the highest confidence level. For reasons of parsimony and model identifiability,
we assume in the present case that the state-response mapping of confidence ratings for
``detect old'' and ``detect new'' states is the same (but see Klauer \& Kellen, 2010).}

\smallskip\noindent
$^2$The non-constrained MPT models can be transformed into equivalent binary MPT models in
a second step which we do not describe, because it is well known (Hu \& Batchelder, 1994).
Furthermore, given the maximum likelihood estimates of 
the parameters of the binary MPT model and an estimate of its Fisher information, 
maximum likelihood estimates of the parameters of the non-constrained MPT model as well as of the
parameters of the original order-constrained MPT model and of their Fisher information matrices (for confidence intervals)
can be obtained via standard methods (Rao, 1973, chap. 6a) using the first derivatives 
of the respective parameter transformation functions that transform these models' parameters into each other although
there is as of yet no user-friendly software to accomplish this.

\newpage
{\renewcommand{\baselinestretch}{1.0}
\small

\begin{table}[htpb]
\captionsetup{textfont=it, labelsep=newline, margin={0 cm, 9 cm}}
\caption{Vertices for the Patterns in Figure 4}
\begin{tabular}{l l c c c c c c c c}\hline
&&\multicolumn{8}{c}{Vertices}    \\
&&\multicolumn{8}{c}{\hrulefill}\\
Pattern&Category&1&2&3&4&5&6&7&8\\                             
\hline
I  &A&1&1/2&1/3&1/4\\
   &B&0&1/2&1/3&1/4\\
   &C&0& 0 &1/3&1/4\\
   &D&0& 0 & 0 &1/4\\    
II &A&1& 0 & 0 &1/4\\
   &B&0& 1 & 0 &1/4\\
   &C&0& 0 & 1 &1/4\\
   &D&0& 0 & 0 &1/4\\ 
III&A&1& 0 &1/3&1/4\\
   &B&0& 1 &1/3&1/4\\
   &C&0& 0 &1/3&1/4\\
   &D&0& 0 & 0 &1/4\\         
IV &A&1& 0 &1/2&1/4\\
   &B&0& 1 &1/2&1/4\\
   &C&0& 0 & 0 &1/4\\
   &D&0& 0 & 0 &1/4\\         
 V &A&1&1/2&1/2&1/3&1/4\\
   &B&0&1/2& 0 &1/3&1/4\\
   &C&0& 0 &1/2&1/3&1/4\\
   &D&0& 0 & 0 &0&1/4\\  
 VI&A&1&1/2&1/3&1/3&1/4\\
   &B&0&1/2&1/3&1/3&1/4\\
   &C&0& 0 &1/3& 0 &1/4\\
   &D&0& 0 & 0 &1/3&1/4\\  
VII&A&1& 0 &1/3&1/3&1/4\\
   &B&0& 1 &1/3&1/3&1/4\\
   &C&0& 0 &1/3& 0 &1/4\\
   &D&0& 0 & 0 &1/3&1/4\\  
VIII&A&1& 0 &1/3& 0 &1/4\\
    &B&0& 1 &1/3&1/2&1/4\\
    &C&0& 0 &1/3& 0 &1/4\\
    &D&0& 0 & 0 &1/2&1/4\\  
IX&A&1& 1/2 &1/2&1/3&1/3&1/4\\
   &B&0& 1/2 & 0 &1/3&1/3&1/4\\
   &C&0& 0   &1/2&1/3& 0 &1/4\\
   &D&0& 0   & 0 & 0 &1/3&1/4\\  
  X&A&1&1/2&1/2&1/2&1/3&1/3&1/3&1/4\\
   &B&0&1/2& 0 & 0 &1/3&1/3& 0 &1/4\\
   &C&0& 0 &1/2& 0 &1/3& 0 &1/3&1/4\\
   &D&0& 0 & 0 &1/2& 0 &1/3&1/3&1/4\\  
\hline
\end{tabular}	
	\label{tab:tab1}
\begin{flushleft}
\end{flushleft}
\end{table}
}
\clearpage

\begin{landscape}
\begin{table}[htpb]
\captionsetup{textfont=it, labelsep=newline, margin={0 cm, 6 cm}}
\caption{Non-Redundant Parameterizations of Mixture Weights for the Patterns in Figure 4}
\begin{tabular}{l c   c c c c}\hline
&\multicolumn{5}{c}{Patterns} \\
Ver-&\multicolumn{5}{c}{\hrulefill}     \\
tices&V&VI&VIII&IX&X\\                           
\hline
1&$(1-\theta_1)(1-\theta_2)(1-\theta_3)$&$\theta_1$&$\theta_1$&$(1-\theta_1)(1-\theta_2)(1-\theta_3)$&$(1-\theta_1)(1-\theta_2)(1-\theta_3)$\\
2&$\theta_1(1-\theta_2)(1-\theta_3)$&$(1-\theta_1)(1-\theta_2)(1-\theta_3)$&$(1-\theta_1)(1-\theta_2)(1-\theta_3) $ &$\theta_1(1-\theta_2)(1-\theta_3)$&$\theta_1(1-\theta_2)(1-\theta_3)$\\
3&$(1-\theta_1)\theta_2(1-\theta_3)$&$(1-\theta_1)\theta_2(1-\theta_3)$& $(1-\theta_1)\theta_2(1-\theta_3)$ &$(1-\theta_1)\theta_2(1-\theta_3)$&$(1-\theta_1)\theta_2(1-\theta_3)$\\
4&$\theta_1\theta_2(1-\theta_3)$&$(1-\theta_1)(1-\theta_2)\theta_3$&$(1-\theta_1)(1-\theta_2)\theta_3   $&$\theta_1\theta_2(1-\theta_3)$&$(1-\theta_1)(1-\theta_2)\theta_3$\\
5&$\theta_3$&$(1-\theta_1)\theta_2\theta_3$&$(1-\theta_1)\theta_2\theta_3$& $(1-\theta_2)\theta_3 $    &$\theta_1\theta_2(1-\theta_3)$\\
6&&&&$\theta_2\theta_3$&$\theta_1(1-\theta_2)\theta_3$\\
7&&&&&$(1-\theta_1)\theta_2\theta_3$\\
8&&&&&$\theta_1\theta_2\theta_3$\\
\hline
\end{tabular}	
	\label{tab:tab2}
\begin{flushleft}
{\em Note.} Patterns I to IV employ four mixture coefficients which are trivial to parameterize
with three non-redundant parameters $\theta_i$ with $0\le\theta_i\le 1$, $i=1,\ldots, 3$.
We believe that no parameterization exists for pattern VII that exhaustively represents
all probability distributions with the order constraints using three non-redundant parameters, 
but did not find a proof for the non-existence of such a parameterization. We used numerical
methods to ascertain that the parameterizations shown exhaust the space of probability distributions
with the appropriate order constraints for all practical purposes (see online supplemental materials).
\end{flushleft}\emph{}
\end{table}
\end{landscape}
\clearpage

\newpage
\pagestyle{fancy}
\begin{landscape}
\begin{figure}[htbp]
 \centering
		\includegraphics[width=1.4\textwidth]{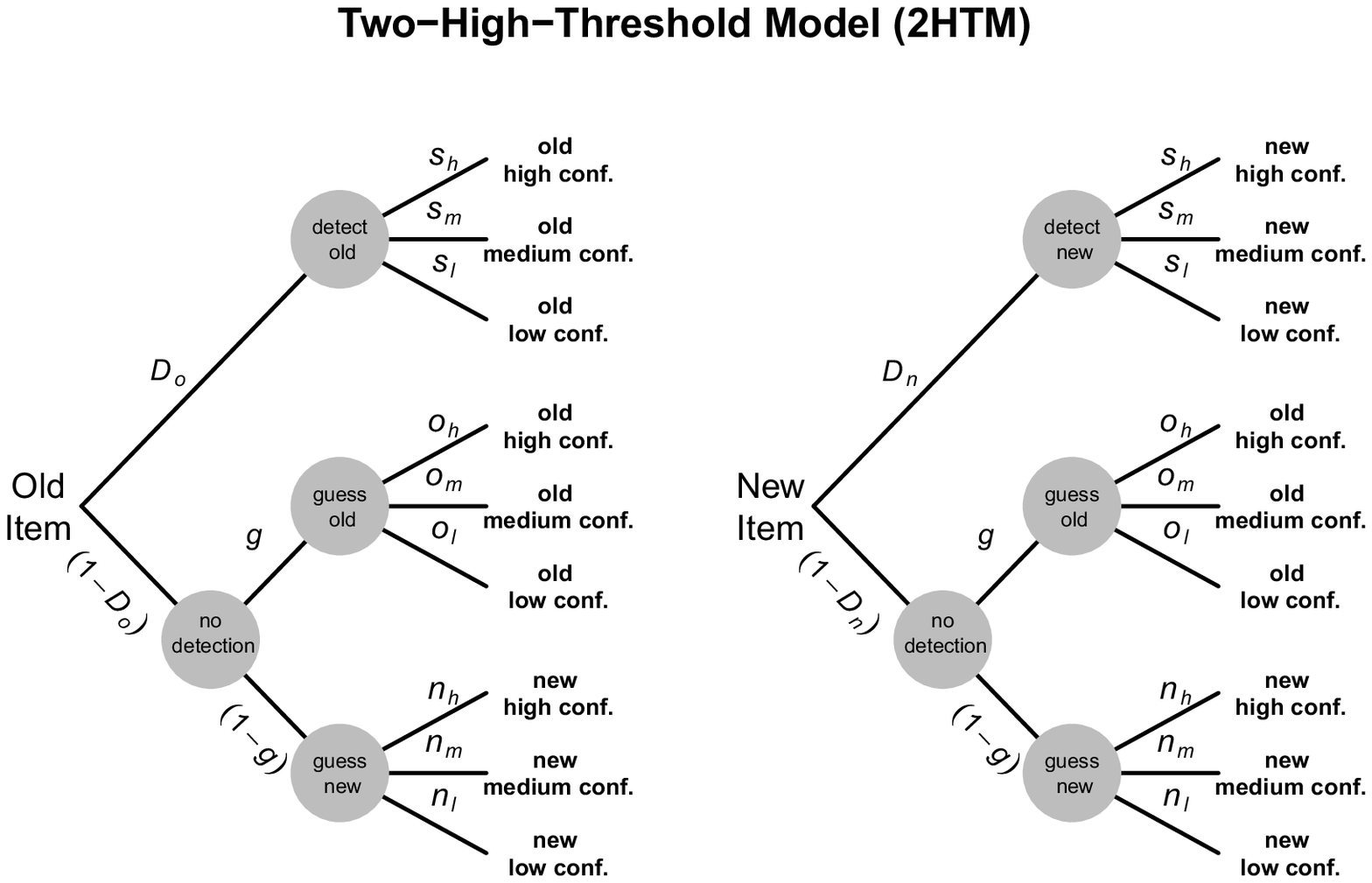}
	\label{fig:Figure1}
\captionsetup{labelfont=it, labelsep=period}
	\caption{The 2HTM model for confidence ratings. Response categories are shown in the middle; the tree to the left models processing
	for old items; the tree to the right for new items.}
\end{figure}
\end{landscape}
\clearpage

\begin{figure}[htbp]
	\centering
		\includegraphics[height= \textheight]{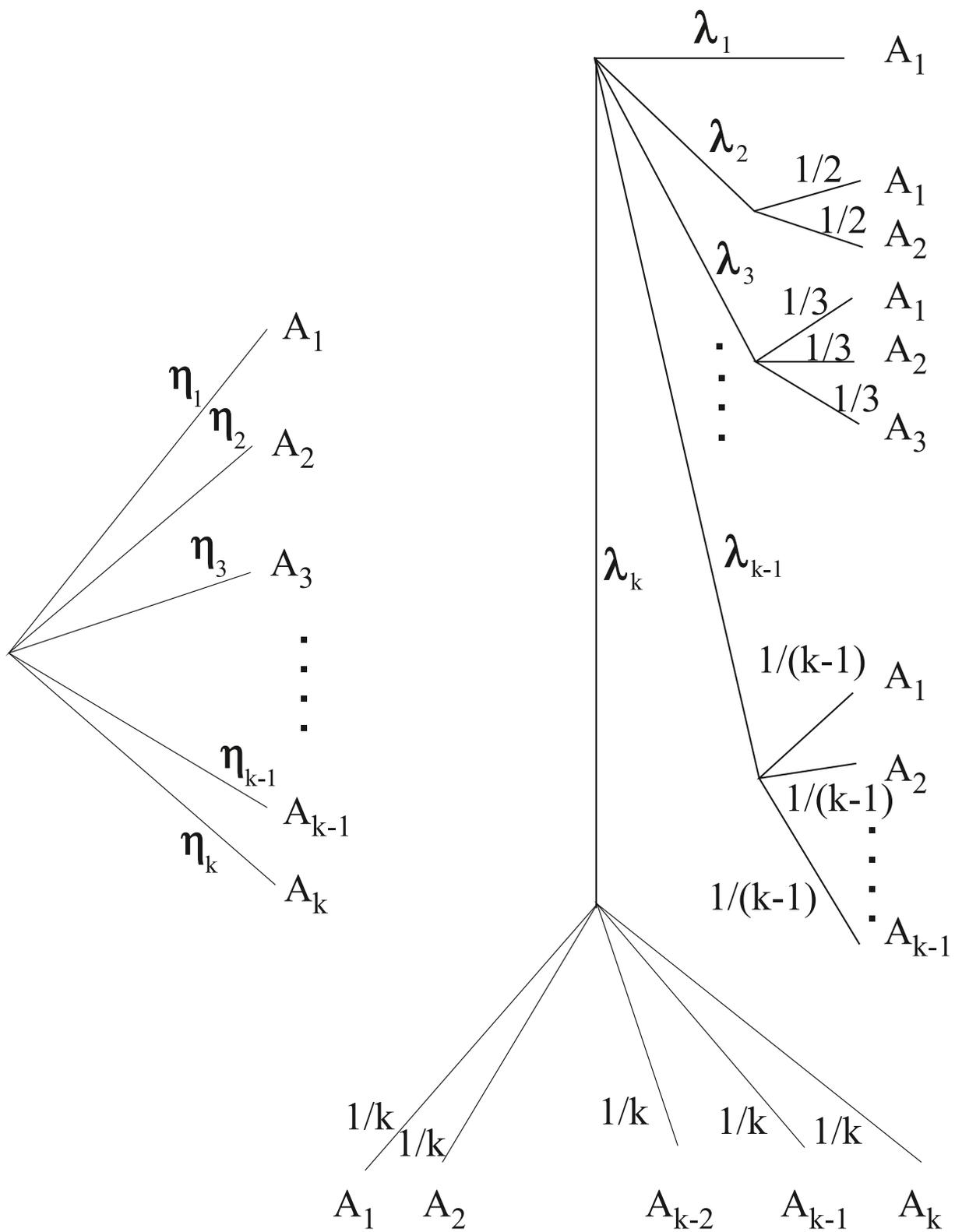}
	\label{fig:Figure2}
\captionsetup{labelfont=it, labelsep=period}
	\caption{The tree to the right is a statistically equivalent reparameterization of the tree to the left
	for ordered parameters $\eta_1\ge \eta_2\ge \ldots \ge \eta_{k-1} \ge \eta_k$.}
\end{figure}

\clearpage

\begin{figure}[htbp]
	\centering
		\includegraphics[width =\textwidth]{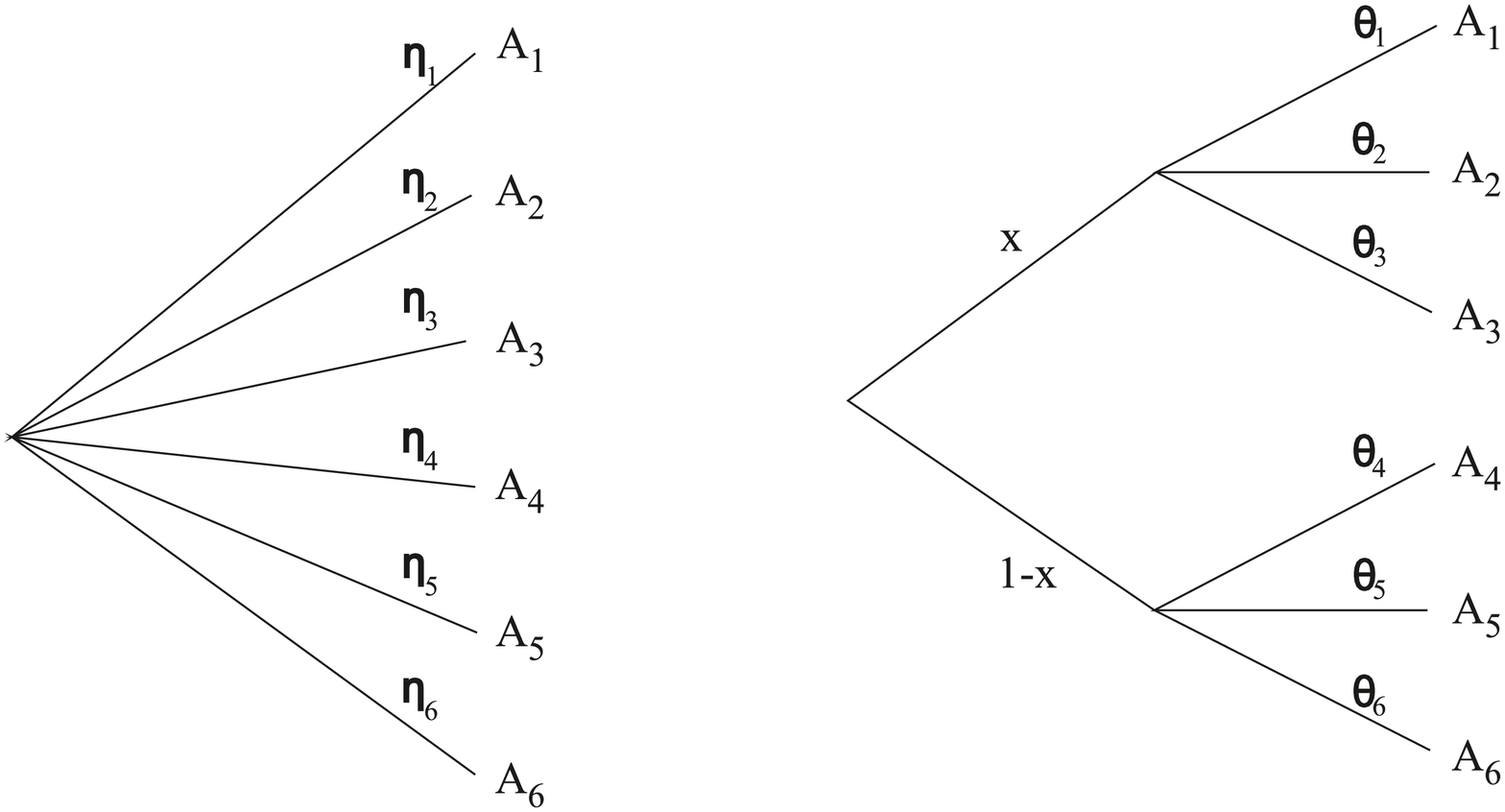}
	\label{fig:Figure3}
\captionsetup{labelfont=it, labelsep=period}
	\caption{The tree to the right is a statistically equivalent reparameterization
	of the tree to the left, if $\sum_i \eta_i =\sum_{i=1}^3\theta_i = \sum_{i=4}^6\theta_i=1$.}
\end{figure}

\begin{landscape}
\begin{figure}[htbp]
	\centering
		\includegraphics[width =1.4\textwidth]{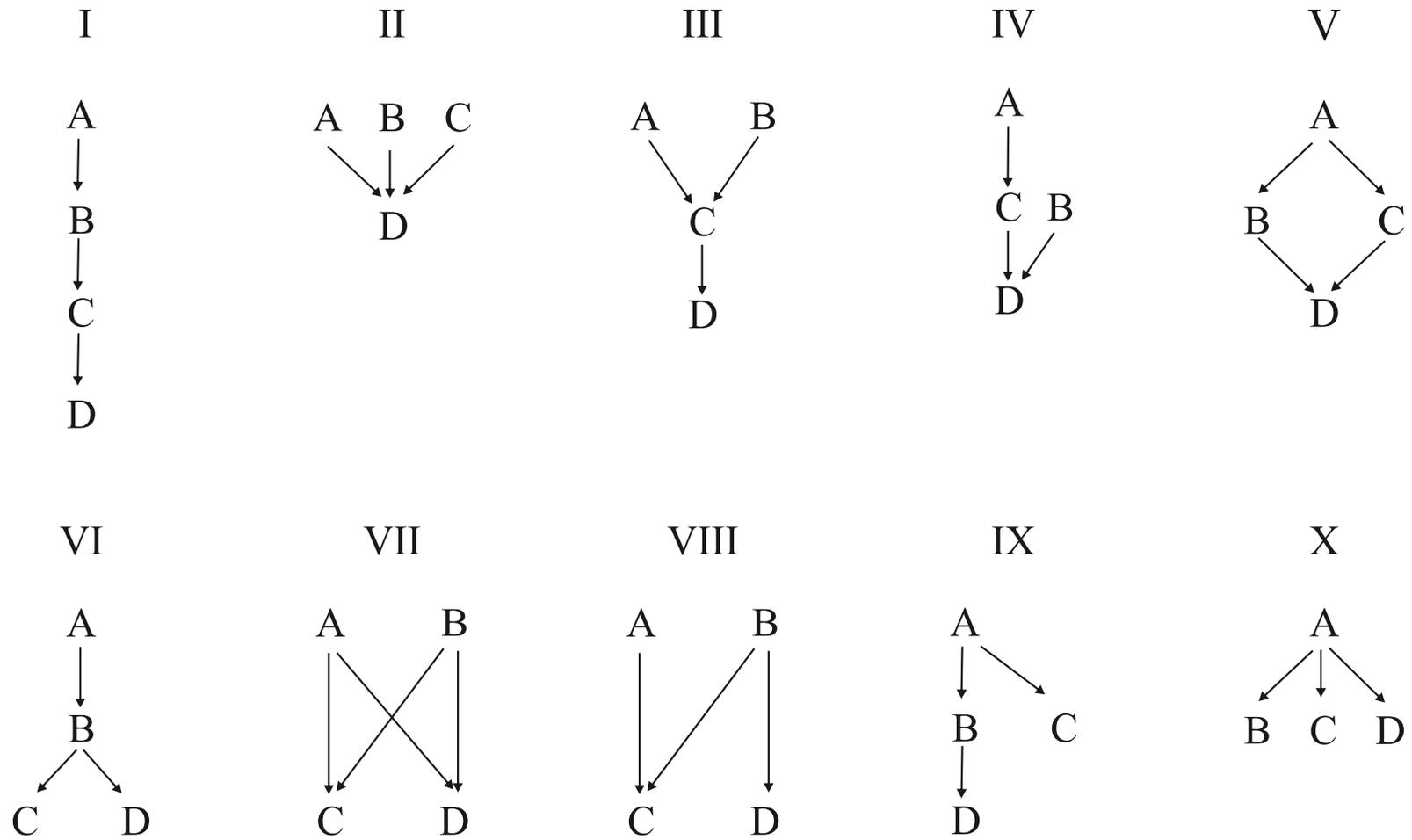}
	\label{fig:Figure4}
\captionsetup{labelfont=it, labelsep=period}
	\caption{Possible orders with four categories A, B, C, and D. Relations implied by transitivity are not shown.}
\end{figure}
\end{landscape}
\clearpage

\clearpage

\end{flushleft}
\end{document}